\documentclass[manuscript, letterpaper]{aastex6}

\hypersetup{colorlinks = false}
\makeatletter 
\long\def\frontmatter@title@above{
\vspace*{-\headsep}\vspace*{\headheight}
\noindent\footnotesize
{\noindent\footnotesize\textsc{\@journalinfo}}\par
{\noindent\scriptsize Preprint typeset using \LaTeX\ style AASTeX6
with modifications by DWH and DFM.
}\par\vspace*{-\baselineskip}\vspace*{0.625in}
}%
\long\def\frontmatter@abstractheading{%
\makeaffils
  \vspace*{-\baselineskip}\vspace*{1.5pt}
  \vspace*{0.13189in}
 \begingroup
  \centering
  \abstractname
  \vskip 1mm
  \par
 \endgroup
 \everypar{\rightskip=0.0in\leftskip=\rightskip}\par
}%
\def\frontmatter@keys@format{\vspace*{0.5mm}%
  \settowidth{\keys@width}{\normalsize\@keys@name}%
  \rightskip=0.0in\leftskip=\rightskip\parindent=0pt%
    \hangindent=\keys@width\hangafter=1\normalsize\raggedright}%
\def\twodigits#1{\ifnum#1<10 0\fi\the#1}
\def\mydate{\leavevmode\hbox{\the\year-\twodigits\month-\twodigits\day}}
\makeatother

\makeatletter
\let\origsection\section
\renewcommand\section{\@ifstar{\starsection}{\nostarsection}}
\newcommand\nostarsection[1]{\sectionprelude\origsection{#1}}
\newcommand\starsection[1]{\sectionprelude\origsection*{#1}}
\newcommand\sectionprelude{\vspace{1em}}
\let\origsubsection\subsection
\renewcommand\subsection{\@ifstar{\starsubsection}{\nostarsubsection}}
\newcommand\nostarsubsection[1]{\subsectionprelude\origsubsection{#1}}
\newcommand\starsubsection[1]{\subsectionprelude\origsubsection*{#1}}
\newcommand\subsectionprelude{\vspace{1em}}
\makeatother

\sloppypar

\definecolor{cbblue}{HTML}{3182bd}
\usepackage{microtype}  
\usepackage{amsmath}
\usepackage{tikz}
\hypersetup{backref,breaklinks,colorlinks,urlcolor=cbblue,linkcolor=cbblue,citecolor=black}
\graphicspath{{figures/}}

\newcommand{\project}[1]{\textsl{#1}}
\newcommand{\acronym}[1]{{\small{#1}}}
\newcommand{\gaia}{\project{Gaia}}
\newcommand{\rave}{\project{\acronym{RAVE}}}

\newcommand{\tmass}{\project{\acronym{2MASS}}}
\newcommand{\documentname}{Article}
\newcommand{\sectionname}{Section}
\newcommand{\figname}{Figure}
\newcommand{\eqname}{Equation}
\newcommand{\dr}{\acronym{DR1}}
\newcommand{\tgas}{\acronym{TGAS}}

\newcommand{\given}{\,|\,}
\newcommand{\normal}{{\mathcal{N}}}
\newcommand{\dd}{\mathrm{d}}
\newcommand{\transp}[1]{{#1}^{\!\mathsf{T}}}
\newcommand{\inv}[1]{{#1}^{-1}}
\newcommand{\bs}[1]{\boldsymbol{#1}}

\newcommand{\propm}{\bs{\mu}}
\newcommand{\mat}[1]{\mathbf{#1}}
\renewcommand{\vec}[1]{\bs{#1}}
\newcommand{\kms}{\ensuremath{\rm km~s^{-1}}}
\newcommand{\msun}{{\rm M}_\odot}
\newcommand{\data}{\mathrm{data}}
\newcommand{\snr}{[S/N]_\varpi}
\newcommand{\eye}{\mathbb{I}}
\newcommand{\absdvtan}{\ensuremath{|\Delta\vec v_\mathrm{t}|}}

\begin{document}\sloppy\sloppypar\raggedbottom\frenchspacing 

\title{Comoving stars in \textsl{Gaia DR1}:\\
  An abundance of very wide separation comoving pairs}
\author{Semyeong Oh\altaffilmark{\pu,\lead},
        Adrian M. Price-Whelan\altaffilmark{\pu},
        David W. Hogg\altaffilmark{\ccpp,\mpia,\cca},
        Timothy D. Morton\altaffilmark{\pu},
        David N. Spergel\altaffilmark{\pu,\cca}
}

\newcommand{\pu}{1}
\newcommand{\lead}{2}
\newcommand{\ccpp}{3}
\newcommand{\mpia}{4}
\newcommand{\cca}{5}

\altaffiltext{\pu}{Department of Astrophysical Sciences,
                   Princeton University, Princeton, NJ 08544, USA}
\altaffiltext{\lead}{To whom correspondence should be addressed:
                     \texttt{semyeong@astro.princeton.edu}}
\altaffiltext{\ccpp}{Center for Cosmology and Particle Physics,
                     Department of Physics,
                     New York University, 4 Washington Place,
                     New York, NY 10003, USA}
\altaffiltext{\mpia}{Max-Planck-Institut f\"ur Astronomie,
                     K\"onigstuhl 17, D-69117 Heidelberg, Germany}
\altaffiltext{\cca}{Center for Computational Astrophysics, Flatiron Institute,
                    162 5th Ave, New York, NY 10003, USA}

\begin{abstract}
The primary sample of the \gaia\ Data Release 1 is the
\textsl{Tycho-Gaia Astrometric Solution} (\tgas): $\approx$ 2 million Tycho-2
sources with improved parallaxes and proper motions relative to the initial
catalog.
This increased astrometric precision presents an opportunity to find new binary
stars and moving groups.
We search for high-confidence comoving pairs of stars in \tgas\ by identifying
pairs of stars consistent with having the same 3D velocity using a marginalized
likelihood ratio test to discriminate candidate comoving pairs from the field
population.
Although we perform some visualizations using (bias-corrected) inverse parallax
as a point estimate of distance, the likelihood ratio is computed with a
probabilistic model that includes the covariances of parallax and proper motions
and marginalizes the (unknown) true distances and 3D velocities of the stars.
We find 13,085 comoving star pairs among 10,606 unique stars with separations
as large as 10 pc (our search limit).
Some of these pairs form larger groups through mutual comoving neighbors:
many of these pair networks correspond to known open clusters and OB
associations, but we also report the discovery of several new comoving groups.
Most surprisingly, we find a large number of very wide ($>1$ pc) separation
comoving star pairs, the number of which increases with increasing separation
and cannot be explained purely by false-positive contamination.
Our key result is a catalog of high-confidence comoving pairs of stars in
\tgas.
We discuss the utility of this catalog for making dynamical inferences about the
Galaxy, testing stellar atmosphere models, and validating chemical abundance
measurements.
\end{abstract}


\section{Introduction} \label{sec:intro}

Stars that are roughly co-located and moving with similar space velocities
(``comoving stars'') are of special interest in many branches of astrophysics.

At small separations ($0.001$--$1$~pc), they are wide binaries (and
multiples) that are either weakly gravitationally bound or
slowly separating.
Because they have low binding energies, a sample of wide binaries is valuable
for investigating the Galactic dynamical environment.
These systems must have both survived their dynamic birth environment and
avoided tidal destruction along their orbit.
Thus, the statistical properties of wide binaries provide a window into both star formation
processes (e.g., \citealt{Parker:2009aa}) and Galactic dynamics (\citealt{Heggie:1975aa}),
including Galactic tides and other massive perturbers such as molecular clouds
and MACHOs \citep{Weinberg:1987aa,Jiang:2010aa,Yoo:2004aa,Allen:2014aa}.

Wide binaries are also good test beds for stellar models and age indicators:
the constituent stars were likely born at the same time with the same chemical
compositions, but evolved independently because of their wide separation.
These pairs are therefore useful for validating gyrochronology relations
(e.g., \citealt{Chaname:2012}) and may be valuable for testing consistency
between stellar atmosphere models.
Finally, calibration of stellar parameters of low-mass stars (e.g., M dwarfs),
which dominate the stellar content of the Galaxy by number, can benefit from
a larger sample of widely separated binaries containing a low-mass star and a
much brighter F/G/K star whose stellar parameters are easier to measure
\citep[e.g.,][]{Rojas-Ayala:2012aa}.

At larger separations ($\gtrsim 1$~pc), comoving stars are likely members of
(potentially dissolving) moving groups, associations, and star clusters or
disrupted wide binaries.
The origin of moving groups is still under active debate (e.g., \citealt{Bovy:2010aa}):
are they remnants of a coeval
star formation event with similar chemical composition? Or are they formed by
dynamical effects of nonaxisymmetric features of the Galaxy such as spirals
and bars? With the recent advances in measuring chemical abundances of
a large volume of stars using high- and low- resolution spectroscopy
(e.g., \citealt{Steinmetz:2006aa,Majewski:2015aa,Gilmore:2012aa} to name a few),
we can now start to explore these questions
with unprecedented statistics and in unexplored detail.
The dynamics of cluster dissolution provides important clues to understanding
the star formation history and the dynamical evolution of the Milky Way.
In the halo, we know of more than 20 disrupting globular clusters and dwarf galaxies
(``stellar streams''; see, e.g., \citealt{Grillmair:2016} for a summary of known streams).
These tidal streams are modeled to infer the parameters of the Galactic
potential (e.g., \citealt{Kupper:2015}).
Similar processes are at work with star clusters in the disk.
However, the dynamical time is much shorter, and the dynamics will be much more
complex because of the existence of other perturbers in the disk.

To date, thousands of candidate comoving star pairs have been identified by
searching for stars with common proper motions
(\citealt{Poveda:1994aa,Allen:2000aa,Chaname:2004aa,
Lepine:2007aa,Shaya:2011aa,Alonso-Floriano:2015aa}).
Here, we use the recent first data release of \gaia\ which includes precise distances,
enabling us to ask whether two stars share the same \emph{physical (3D) velocity} rather than
just the projections in the proper motion space.

This paper proceeds as follows:
In \sectionname~\ref{sec:data}, we briefly describe the data set used in
this work.
In \sectionname~\ref{sec:methods}, we develop a statistical method to identify
high-confidence comoving pairs in this catalog.
In \sectionname~\ref{sec:results}, we present and discuss our resulting catalog
of comoving pairs.
We summarize in \sectionname~\ref{sec:summary}.


\section{Data} \label{sec:data}

The primary data set used in this \documentname\ is the Tycho-Gaia Astrometric
Solution (\tgas), released as a part of Data Release 1 (\dr) of the Gaia mission
\citep{Gaia-Collaboration:2016aa,Lindegren:2016aa}.
The \tgas\ contains astrometric measurements (sky position,
parallax, and proper motions) and associated covariance matrices for a large
fraction of the \project{Tycho-2} catalog \citep{2000A&A...355L..27H} with median
astrometric precision comparable to that of the \project{Hipparcos} catalog
\citep[$\approx 0.3~{\rm mas}$;][]{2007ASSL..350.....V}. In terms of parallax
signal-to-noise ($\snr = \varpi/\sigma_\varpi$), the \tgas\ catalog contains
42385 high-precision stars with $\snr > 32$.

We construct an initial sample of star pairs to search for comoving pairs as
follows.
We first apply a global parallax signal-to-noise cut, $\snr > 8$,  to the \tgas,
which leaves 619,618 stars.
Then, for each star we establish an initial sample of possible
comoving partners by selecting all other stars that satisfy two criteria:
separation less than 10~pc and difference
in (point-estimate) tangential velocity less than $\absdvtan < 10$~\kms.
We ultimately build a statistical model that incorporates the covariances of the
data, but for these initial cuts and for visualizations we use a point-estimate
of the distance by applying a correction for the Lutz-Kelker bias
(\citealt{Lutz:1973aa}):
\begin{equation}
  \hat{r} = 1000 \, \left[\frac{\varpi}{2} \,
    \left(1 + \sqrt{1 - \frac{16}{\snr^2}} \right) \right]^{-1} \, {\rm pc}
    \label{eq:dist}
\end{equation}
where $\varpi$ is the parallax in mas.
An estimate for the difference in tangential velocity between two stars is,
then,
\begin{equation}
  \absdvtan = |\hat{r}_1 \vec\mu_1 - \hat{r}_2 \vec\mu_2|
\end{equation}
where $\vec\mu = (\mu_{\alpha^*}, \mu_\delta)$.
\footnote{$\mu_{\alpha^*}$ is the proper motion component
in the right ascension direction, $\mu_{\alpha^*} = \mu_\alpha \cos\delta$}

\begin{figure}[htbp]
  \begin{center}
    \includegraphics[width=\linewidth]{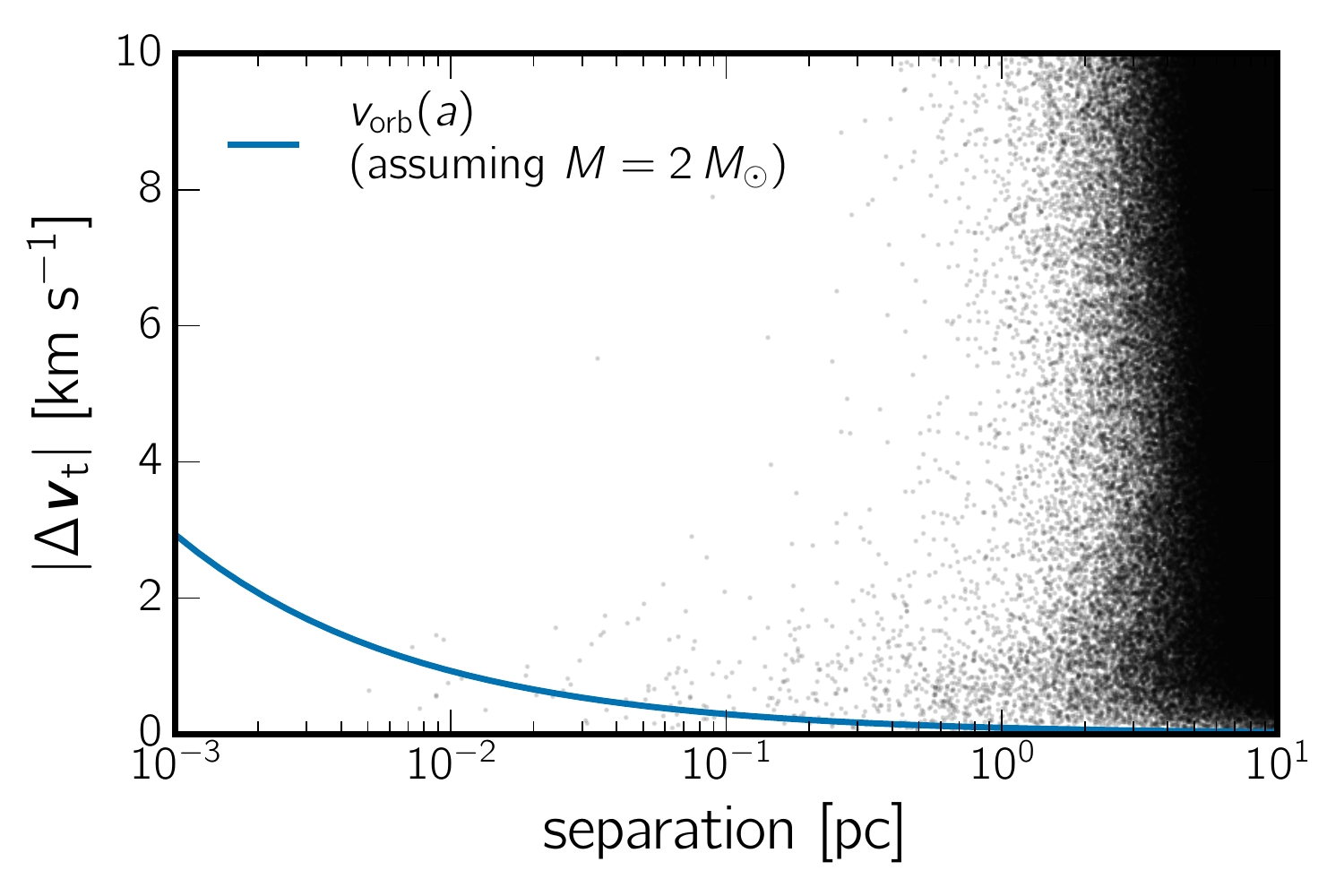}
  \end{center}
  \caption{%
    Point estimates of tangential velocity and physical separation computed for
    all 271,232 unique pairs in the initial sample of pairs (black points).
    For this sample, we consider stars
    with separation $< 10$~pc and $\absdvtan < 10$~\kms\
    computed relative to every other star in \tgas.
    The blue solid line shows the magnitude of the 3D
    orbital velocity as a function of semi-major axis for a $2~M_\odot$ binary system.
    Note the stream of points that starts at small separation ($\lesssim 0.01$~pc),
    small $\absdvtan$ ($\lesssim 2$~\kms),
    but climbs to larger $\absdvtan$ at $|\Delta \vec{x}|\gtrsim 1$~pc, which
    eventually merges with random pairs of field stars dominating in the upper right corner
    (see \sectionname~\ref{sec:data} for details).
    \label{fig:dv-sep}}
\end{figure}

Figure~\ref{fig:dv-sep} shows \absdvtan\ against the physical separation
for the resulting 271,232 unique pairs in the initial sample.
A few key observations can be made:

\begin{itemize}
  \item At small separations ($<1$~pc), there is a population of pairs with
    very small tangential velocity difference ($<2$~km/s). Given that these
    pairs are very close in both 3D position and tangential velocities, it is highly
    probable that they are actually comoving wide binaries.

  \item  A sample of comoving stars also include stars that are part of, e.g.,
    OB associations, moving groups, and open clusters.
    These astrophysical objects may be detected as a network of comoving pairs,
    sharing some mutual comoving neighbors.
    As the pair separation increases, the nature of comoving pairs
    will change from binaries to those related to these larger objects,
    which generally subtend a larger angle in sky.
    Since the proper motions of two stars with the same 3D velocity
    are projections of this velocity onto the celestial sphere at
    two different viewing angles,
    the larger the difference in viewing angles is, the larger the difference in tangential
    velocities will be.
    Due to this projection effect, a population of genuine comoving pairs
    will extend to larger \absdvtan\ at larger separation.
    This indeed can be seen in Figure~\ref{fig:dv-sep} as an over-density
    in the lower right corner that gets thinner as \absdvtan\ increases.

  \item Finally, there is a population of ``random'' pairs of field stars
    that are not comoving, but still have $\absdvtan < 10$~\kms\
    by chance.
    As \absdvtan\ increases, this population will dominate.
    Figure~\ref{fig:dv-sep} shows that there is an overlap between
    genuine comoving pairs and ``random'' pairs.
\end{itemize}

In the following section, we construct a statistical model that propagates
the non-trivial uncertainties in the data to our beliefs about the likelihood
that a given pair of stars is comoving.

\section{Methods} \label{sec:methods}

The abundance of pairs of stars with small velocity difference in
\figname~\ref{fig:dv-sep} suggests that there are a
significant number of comoving pairs in the \tgas\ data at a range
of separations.
Here, we develop a method to select high-confidence comoving
pairs that properly incorporates the uncertainties associated with the
\gaia\ data. We make the following assumptions in order to construct a
statistical model (a likelihood function with explicit priors on our
parameters):
\begin{itemize}
  \item We assume that the uncertainties in the data---parallax, $\varpi$, and
    two proper motion components, $\propm = \transp{(\mu_{\alpha^*}\, \mu_\delta)}$
    ---are Gaussian with known covariances
    $\mat{C}$. The values of covariances are provided as part of the \gaia\ data
    \citep{Lindegren:2012aa,Lindegren:2016aa}.
  \item We assume that the 3D velocities of stars in a given pair
    $(\vec{v}_i, \vec{v}_j)$ (relative to the solar system
    barycenter) are either (1) the same with a small (Gaussian) dispersion $s$
    or (2) independent.
    In both cases, velocity is drawn from
    the velocity prior $p(\vec{v})$.
\end{itemize}

Under these assumptions, the likelihood of a proper motion measurement,
$\bs{\mu}$, for a star with true distance, $r$, and true 3D velocity $\vec{v}$ is
\begin{align}
  L(\vec{\mu} \given \vec{v}, r, s^2) &=
    \left[\det\left(\frac{\tilde{\mat{C}}^{-1}}{2\pi}\right)\right]^{1/2} \,
    \exp \left[ -\frac{1}{2} \transp{\left(\vec{\mu} - \vec{x}_\theta \right)} \,
    \tilde{\mat{C}}^{-1} \,
    \left(\vec{\mu} - \vec{x}_\theta \right) \right] \label{eq:likefn} \\
  \vec{x}_\theta &= r^{-1} \, \vec{v}_t
\end{align}
where the tangential velocity $\vec{v}_t = (\begin{array}[t]{c c} v_\alpha & v_\delta\end{array})^\mathsf{T}$
is related to the 3D velocity
$\vec{v}$ by projection matrix $\mat{M}$ at the star's sky position
$(\alpha, \delta)$
\begin{align}
  \vec{v}_t &= \mat{M}\,\vec{v} \\
  & = \left(
      \begin{array}{c c c}
        -\sin\alpha & \cos\alpha & 0 \\
        -\sin\delta \, \cos\alpha & -\sin\delta \, \sin\alpha & \cos\delta
      \end{array}
    \right) \,
    \left(\begin{array}{c} v_x \\ v_y \\ v_z \end{array}\right) \label{eq:transformation}
\end{align}
and the modified covariance matrix $\tilde{\mat{C}}$ is
\begin{equation}
  \tilde{\mat{C}} = \mat{C} + (s/r)^2 \, \eye
\end{equation}
where $\eye$ is the identity matrix.
The parameter $s$ is added to allow for small tolerance in velocities which we discuss below.

For a given pair, we compute the fully marginalized likelihood (FML) for the hypotheses (1) and (2),
$\mathcal{L}_1$ and $\mathcal{L}_2$.
We use the FML ratio $\mathcal{L}_1/\mathcal{L}_2$ as the scalar quantity to select
candidate comoving pairs, as described in $\sectionname~\ref{sub:selection}$ in
more detail.
To compute these FMLs, the likelihood functions for each star in a pair, $L_i,
L_j$, are marginalized over the (unknown) true distance and 3D velocity for each star in
the pair $(i,j)$.
\begin{align}
  \mathcal{L}_1 &=
    \int \, \dd r_i \, \dd r_j \, \dd^3 \vec{v} \,
    L_i(\vec{\mu_i} \given \vec{v}, r_i, s^2) \,
    L_j(\vec{\mu_j} \given \vec{v}, r_j, s^2) \,
    p(\vec{v}) \, p(r_i \given \varpi_i) \, p(r_j \given \varpi_j) \label{eq:hyp1}\\
  \mathcal{L}_2 &=
    \int \, \dd r_i \, \dd r_j \, \dd^3 \vec{v}_i \, \dd^3 \vec{v}_j \,
    L_i(\vec{\mu_i} \given \vec{v}_i, r_i, s^2) \,
    L_j(\vec{\mu_j} \given \vec{v}_j, r_j, s^2) \,
    p(\vec{v}_i) \, p(\vec{v}_j) \, p(r_i \given \varpi_i) \, p(r_j \given \varpi_j). \label{eq:hyp2}
\end{align}
Here, $p(r_i \given \varpi_i)$ is the posterior distribution of distance given parallax measurement
$\varpi_i$ and its Gaussian error $\sigma_{\varpi,i}$.
Note that the FML for the hypothesis (1) involves integration over one velocity $\vec{v}$ that
generates the likelihoods for both stars, $L_i$ and $L_j$.
The marginalization integral for hypothesis (2) can be split into the product of
two simpler integrals $\mathcal{L}_2 = Q(\vec{\mu_i}, \varpi_i) \, Q(\vec{\mu_j}, \varpi_j)$ where
\begin{equation}
  Q(\vec{\mu}, \varpi) = \int \, \dd r \, \dd^3 \vec{v} \, L(\vec{\mu} \given \vec{v}, r, s^2) \,
  p(\vec{v}) \, p(r \given \varpi)
\end{equation}

If the velocity prior $p(\vec{v})$ is also Gaussian, the integrals over velocity
in both cases can be performed analytically:
We use a mixture of three isotropic, zero-mean Gaussian distributions
\begin{equation}
  p(\vec{v}) = \sum_{m=1}^3 \, w_m \, \mathcal{N}(0, \sigma_{v,m}^2)
  \label{eq:vprior}
\end{equation}
with velocity dispersions $(\sigma_{v,1}, \sigma_{v,2}, \sigma_{v,3}) = (15, 30, 50)
~\kms$ and weights $(w_1,w_2,w_3) = (0.3, 0.55, 0.15)$
meant to encompass young thin disk stars to halo stars.
These numbers are empirically chosen to account for the distribution of
velocities of the \tgas\ stars.
We derive the relevant expressions in Appendix~\ref{sec:appendix}.
After marginalizing over velocity, the likelihood integrands only depend on
distance; we numerically compute the integrals over the true distances of each
star in a pair using Monte Carlo integration with $K$ samples from
the distance posterior.
\begin{equation}
  \int \, \dd r \, \tilde{L}(r) \, p(r \given \varpi) \approx
    \frac{1}{K} \, \sum_k^K \, \tilde{L}(r_k)
\end{equation}
where $\tilde{L}(r)$ is the velocity-marginalized likelihood function.
In order to generate a sample of distances from the distance posterior $p(r_i \given \varpi_i)$,
we need to assume a distance prior. We adopt the uniform density prior
\citep{Bailer-Jones:2015aa} with a maximum distance of 1~kpc.
Through experimentation, we have found that $K=128$ samples are sufficient for
estimating the above integrals for stars with a wide range in parallax
signal-to-noise.

For small-separation binaries, the assumption that the stars having the same 3D velocity for
hypothesis (1) can
break down for high-precision proper motion measurements because of the orbital
velocity (blue solid line in Figure~\ref{fig:dv-sep}).
To account for this, we set $s^2 = \frac{2 \, G \, \msun}{|\vec{x}_i-\vec{x}_j|}$.
Because $s^2$ is much smaller than the velocity dispersions of the velocity prior
$p(\vec{v})$, it has minimal effect on the hypothesis (2) FML.

\section{Results} \label{sec:results}

This section is divided into three parts. First, we discuss and justify a cut of
the likelihood ratio to select candidate comoving pairs.
Second, we present the statistics and properties of our candidate comoving
pairs.
Finally, we describe our catalog of candidate comoving pairs, the main product
of this study.

\subsection{Selecting candidate comoving pairs}
\label{sub:selection}

In this section, we examine the distribution of (log-)likelihood ratios
$\ln \mathcal{L}_1 /\mathcal{L}_2$, and come up with a reasonable cut
for this quantity to select comoving pairs from the initial sample.

\begin{figure*}[htbp]
  \begin{center}
    \includegraphics[width=\textwidth]{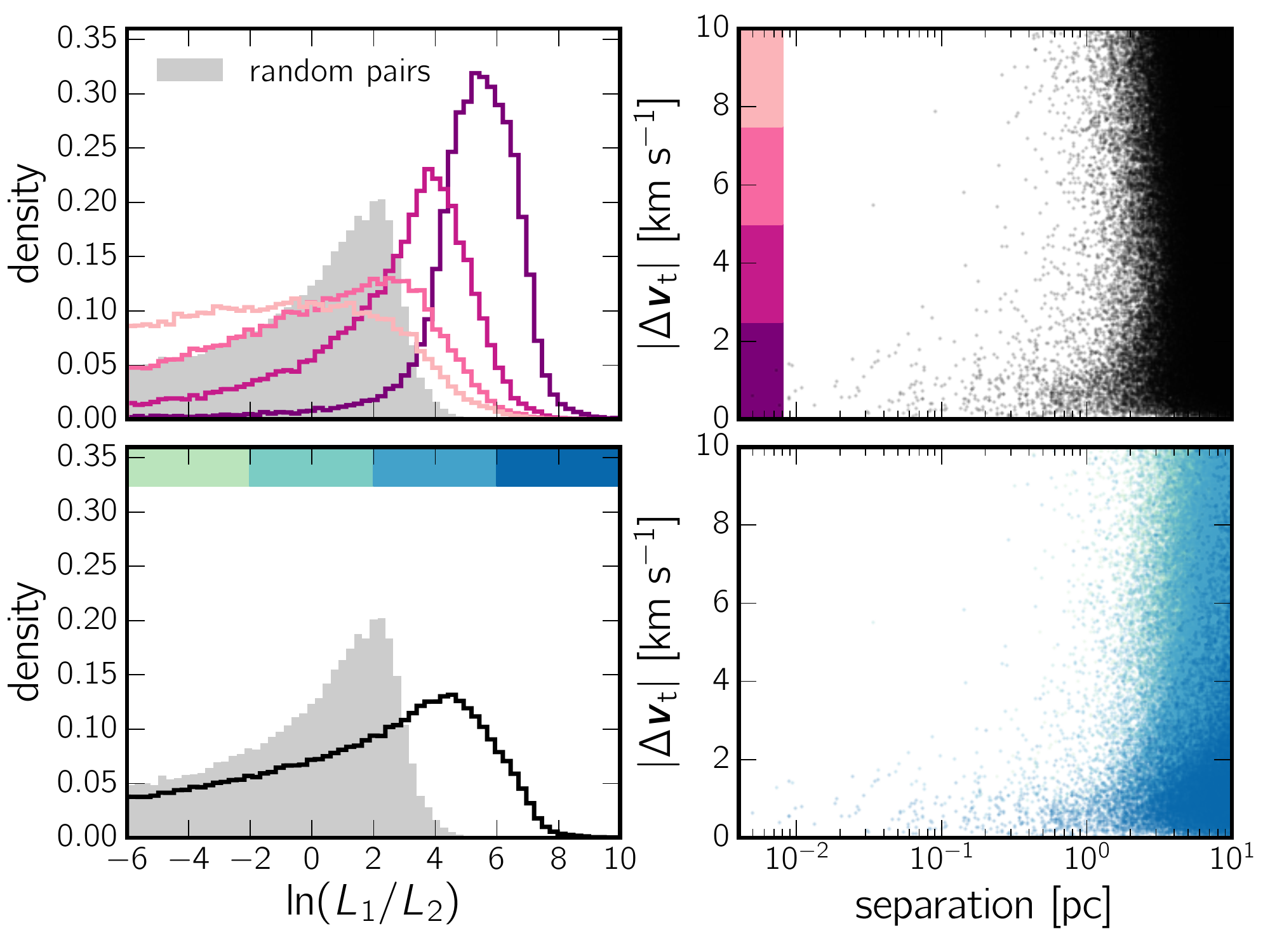}
  \end{center}
  \caption{%
    Justification of the likelihood ratio cut.
    On the top row, we show how the distribution of $\ln \mathcal{L}_1 /\mathcal{L}_2$
    changes (left) in slices of \absdvtan\ (sequentially color-coded in pink on the right panel).
    On the bottom row, we show how the distribution of pairs on \absdvtan\ vs. separation changes (right)
    in slices of $\ln \mathcal{L}_1 /\mathcal{L}_2$ (sequentially color-coded in blue on the left).
    The black dots in the upper right panel and
    the black line in the lower left panel correspond to the entire initial sample of pairs.
    For comparison, we also present the likelihood ratio distribution of random pairs
    of stars in the \tgas\ with parallax $S/N>8$ in gray filled histogram.
    Based on this, we choose $\ln \mathcal{L}_1 /\mathcal{L}_2 > 6$ as our high-confidence
    candidate comoving pairs.
    \label{fig:likelihoodratios}}
    \vspace{1em}
\end{figure*}

Figure~\ref{fig:likelihoodratios} shows the likelihood ratios for all
$\sim 271$k pairs in the initial sample.
As discussed in \sectionname~\ref{sec:data}, we expect a correlation between the likelihood
ratios of pairs, and their distribution on \absdvtan\ vs separation plane.
Specifically, as we sweep through from small \absdvtan\ to large, we expect
the population of pairs to change from genuinely comoving to random.
This becomes clear when we look at the distribution of the likelihood ratios of pairs
in slices of \absdvtan\ (top row of Figure~\ref{fig:likelihoodratios}).
Pairs with $\absdvtan \in (0,2.5)~\kms$ are most likely actual comoving pairs, and
their likelihood ratio distribution is narrowly peaked at $\gtrsim 5$
(darkest pink histogram in the upper left panel of Figure~\ref{fig:likelihoodratios}).
The distribution peaks at lower values and gets broader as \absdvtan\ increases,
and the number of random pairs increasingly dominate.
On the bottom row of Figure~\ref{fig:likelihoodratios}, we show
how the distribution of pairs on \absdvtan\ vs separation plane changes
with decreasing $\ln \mathcal{L}_1 /\mathcal{L}_2$ ratios.
This is in agreement with our discussion in \sectionname~\ref{sec:data}.
Finally, as a test,
we compute the likelihood ratios for 200,000 random pairs of stars with the same
parallax signal-to-noise ratio cut as the initial sample ($S/N > 8$).
Shown as the gray filled histogram in Figure~\ref{fig:likelihoodratios},
this distribution peaks at a much lower value ($\approx 2$),
and is clearly separated from highly probable comoving pairs.

Based on these comparisons, we select candidate comoving pairs with
$\ln \mathcal{L}_1 /\mathcal{L}_2 > 6$.
Out of 271,232 pairs in the initial sample, 13,058 pairs (4.8\%)
satisfy this condition.

\subsection{Statistics and properties of the identified comoving pairs}

\begin{figure}[htbp]
  \begin{center}
    \includegraphics[width=\linewidth]{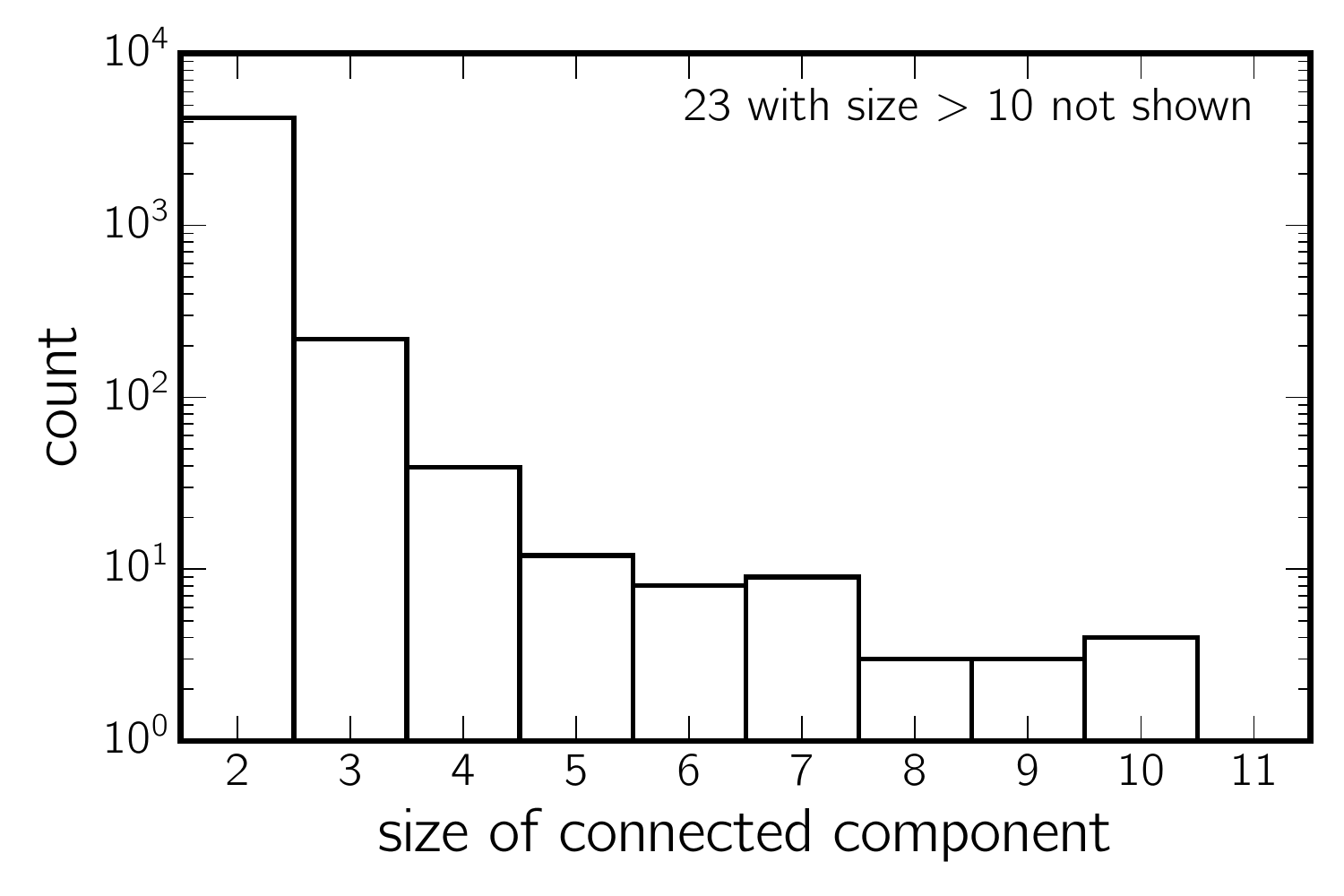}
  \end{center}
  \caption{%
    Histogram of the sizes of connected components.
    \label{fig:hist_ccsize}}
\end{figure}

\begin{figure*}[htbp]
  \begin{center}
    \includegraphics[width=\textwidth]{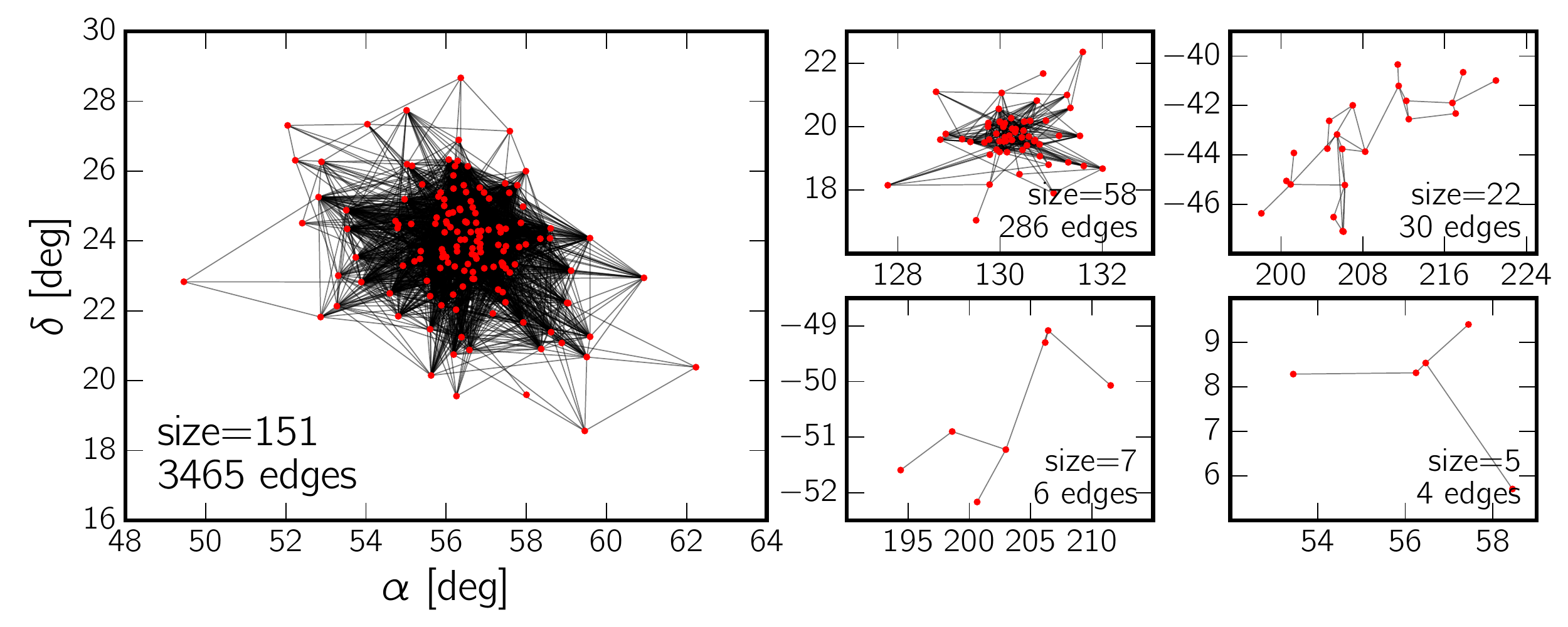}
  \end{center}
  \caption{%
    Visualizations of a few example connected components of comoving pairs of stars.
    Each star (node) is marked as a red circle, and a line (edge) is drawn
    between two stars if they are comoving by our selection criteria
    (see \sectionname~\ref{sub:selection}). On left, we show the largest network
    found in this study corresponding to the Pleiades star cluster.
    On right, we show four examples of connected components with varying sizes.
    The connected component on the upper left panel with a size of 58 corresponds
    to NGC~2632, also known as the Beehive cluster.
    \label{fig:graphviz_examples}}
\end{figure*}

\begin{figure*}[htbp]
  \begin{center}
    \includegraphics[width=\textwidth]{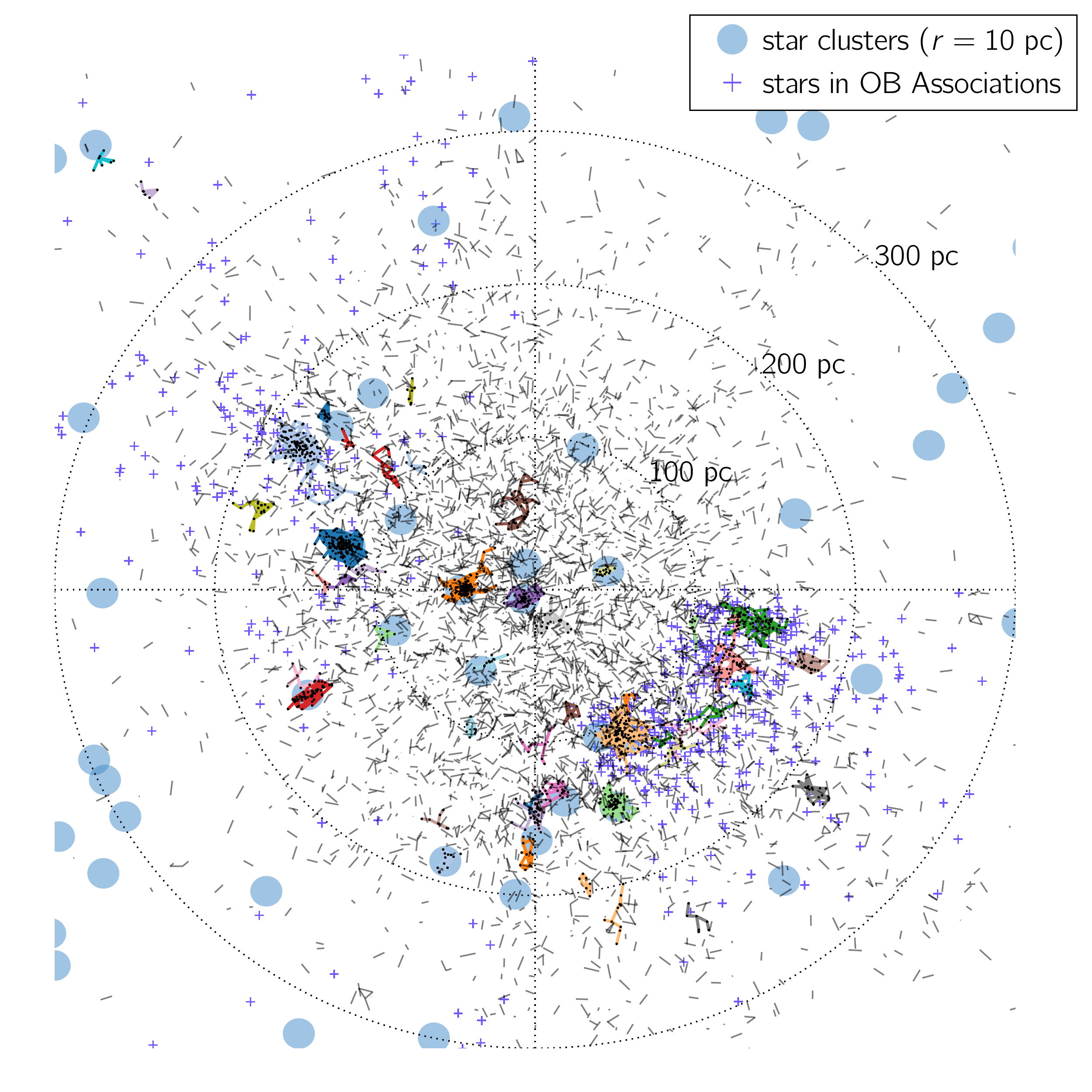}
  \end{center}
  \caption{%
    Panoramic view of comoving pairs of stars around the Sun. This is a cylindrical
    projection of onto the Galactic plane with the Sun at the origin.
    Pairs in connected components of sizes less than 5 are connected by gray lines.
    Pair in connected components of size $\geq 5$ are plotted with a unique random color,
    and their nodes (stars) are highlighted with small black circles.
    We also show the positions of known Milky Way star clusters
    from \citet{Kharchenko:2016aa} as light blue circles of $10$~pc radius,
    and stars in OB associations from \citet{de-Zeeuw:1999aa} as
    blue crosses.
    Some of the larger connected components are clearly associated with known
    clusters, but we also discover quite a few new comoving groups.
    \label{fig:glon_d_pairlines}}
\end{figure*}

Once we have identified candidate comoving pairs from the initial
sample, these pairs form an undirected graph where stars are nodes, and edges
between the nodes exist for comoving pairs of stars.
A star may have multiple comoving neighbors, and two stars may either
be directly or indirectly connected by a sequence of edges (``path'').
We divide the graph into connected components\footnote{A connected component of an undirected graph $G$
is a subgraph of $G$ in which any two nodes are connected to each other by a path.},
and show the distribution of their sizes in Figure~\ref{fig:hist_ccsize}.
The most common are connected components of size 2, which mean mutually exclusive comoving
pairs. However, it is clear that there are many aggregates of comoving stars
discovered by looking for comoving pairs.
These aggregates are likely moving groups, OB associations, or star clusters.
In 13,058 comoving pairs that we identified,
there are 4,555 connected components among 10,606 unique stars.
The maximum size of the connected components is 151.
We show this largest connected component along with four other
examples of varying sizes in Figure~\ref{fig:graphviz_examples}.
The largest connected component corresponds to the Pleiades open cluster (left panel of Figure~\ref{fig:graphviz_examples})
while the upper left panel of the right column of Figure~\ref{fig:graphviz_examples}
is NGC~2632, another known Milky Way open cluster \citep{Kharchenko:2016aa}.

We show the distribution of comoving pairs in galactic longitude and
distance in Figure~\ref{fig:glon_d_pairlines}.
The connection between known comoving structures and
the larger connected components found in this work becomes immediately clear when
we overplot the positions of known Milky Way star clusters \citep{Kharchenko:2016aa},
and stars in OB associations \citep{de-Zeeuw:1999aa}.
Many of the larger connected components are
clearly associated with known star clusters:
Melotte 22 (Pleiades) at $(l,d)\approx (167\arcdeg, 130~\textrm{pc})$,
Melotte 20 at $(l,d)\approx (147\arcdeg, 175~\textrm{pc})$,
Melotte 25 (Hyades) at $(l,d)\approx (180\arcdeg, 50~\textrm{pc})$, and
NGC 2632 (Beehive) at $(l,d)\approx (206\arcdeg, 187~\textrm{pc})$
to name a few.
Clumps of comoving pairs at $(l,d)\approx (300-360\arcdeg, 100-200~\textrm{pc})$
seem to strongly correlate with the locations of OB associations
Upper Scorpius, Upper Centaurus Lupus, and Lower Centaurus Crux
\citep{de-Zeeuw:1999aa}.

However, there are still many new larger connected components that we
discover.
If we define a condition to associate a connected component to a known cluster
as having more than 3 members within 10~pc from the nominal position of
the cluster, for the 61 connected components with sizes larger than 5,
we find that only 10 are associated with a cluster cataloged in
\citet{Kharchenko:2016aa}.
It is also worth noting that the positions of some of the known clusters
are offset from those of the connected components associated with them,
indicating that the \tgas\ data improves the distance estimates of these clusters.
Finally, not all known star clusters are recovered in our search.
This is primarily because of the non-uniform coverage and magnitude limit of the
\tgas\ data.

\begin{figure}[htbp]
  \begin{center}
    \includegraphics[width=0.6\linewidth]{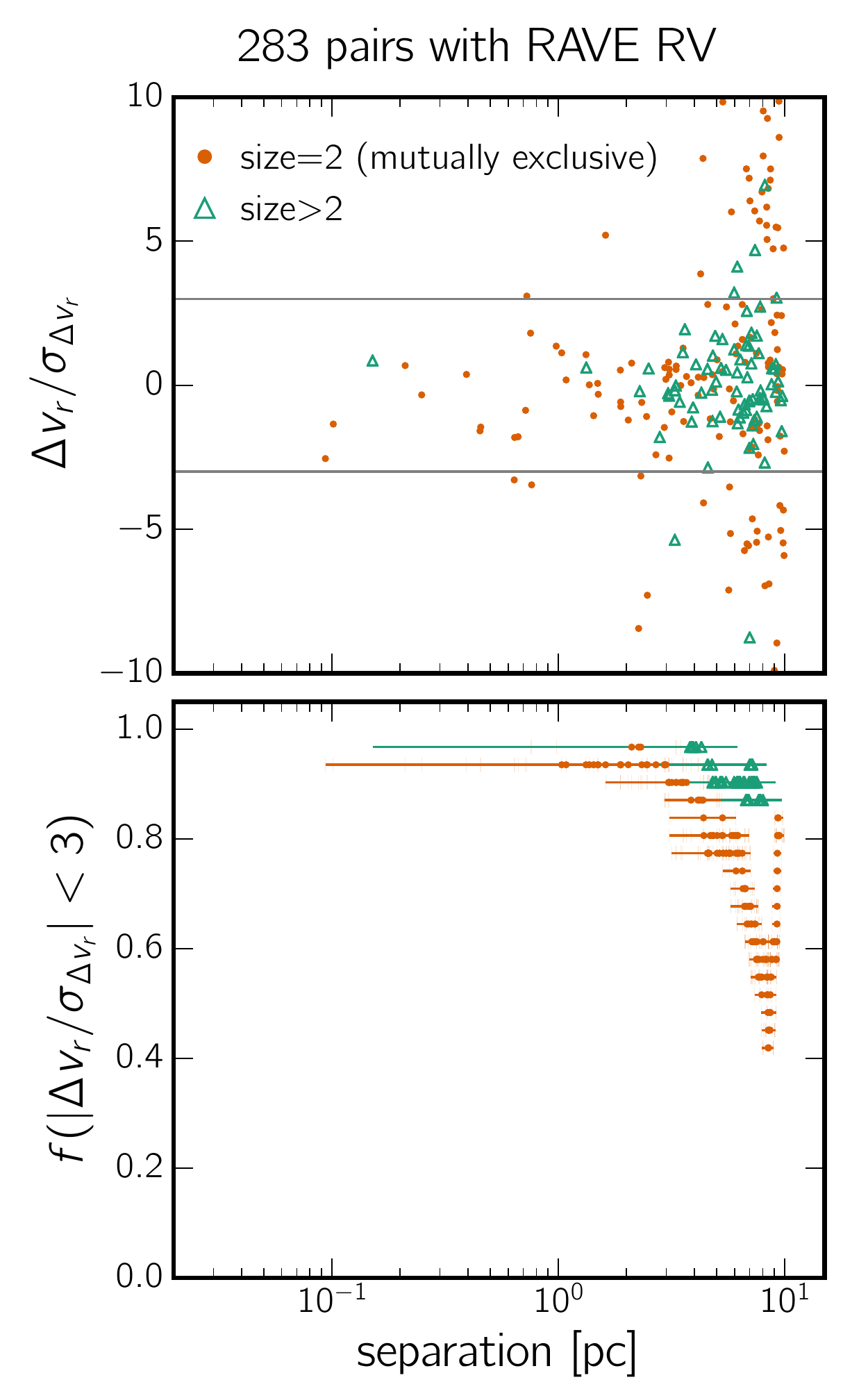}
  \end{center}
  \caption{%
    Validation of candidate comoving pairs using radial velocities from \rave.
    Top: Radial velocity differences, $\Delta v_r$, of 283 pairs with \rave\ measurements.
    $\Delta v_r$ is plotted in units of $\sigma_{\Delta v_r} = \sqrt{\sigma_{v_r,1}^2 + \sigma_{v_r,2}^2}$
    as a function of physical separation.
    We highlight $3\sigma$ limit with two horizontal lines.
    Bottom: Fraction of pairs with $|\Delta v_r/\sigma_{\Delta v_r}| <3$
    as a function of physical separation. We calculate the fraction with a running
    bin containing 31 data points at a time.
    The median, and the minimum and maximum separation of pairs in each bin are
    indicated with a marker, and its errorbars.
    We separate pairs in connected components of size 2 (i.e., mutually exclusively connected)
    from those in larger connected components.
    \label{fig:raverv}}
\end{figure}

Ultimately, any candidate comoving pair found in this work
needs to be verified using radial velocities.
Here, we use 210,368 cross-matches of the \tgas\ with
Radial Velocity Experiment fifth data release (\rave\ DR~5; \citealt{2017AJ....153...75K}) to
assess the false-positive rates of our selection.
We have 283 pairs with both stars matched with \rave.
Figure~\ref{fig:raverv} shows the difference in radial velocities
between the two stars in a pair, $\Delta v_r$, as a function of their physical separation.
We show $\Delta v_r$ in units of $\sigma_{\Delta v_r}$ which we estimate
as the quadrature sum of $\sigma _{v_r}$ for each star.
The fraction of pairs with good agreement in radial velocity decreases with
increasing separation.
This, after all, is not surprising because we are only using
2D velocity information (proper motions) with errors.
However, the contamination becomes significant only at $>1$~pc
(which depends on the local stellar number density and velocity dispersion).
Given the excellent correspondence between
aggregates of comoving pairs (connected components)
and known genuine comoving structures (Figure~\ref{fig:glon_d_pairlines}),
we may expect that pairs in these larger connected components,
which will often have separations $>1$~pc,
to have less contamination from false-positives.
We divide the comoving pairs into those
mutually exclusively connected (i.e., in a connected component of
size 2), and those in a larger group.
We indeed find that many pairs in larger groups are at $>1$~pc, yet
the fraction of pairs that have identical radial velocities within $3\sigma$ is
higher than the mutually exclusive pairs, and remains high ($>80\%$) to $\sim 10$~pc.
Finally, we note that the false-positive rate for mutually exclusive pairs
with large angular separation may have been over-estimated due to projection.

\begin{figure*}[htbp]
  \begin{center}
    \includegraphics[width=\textwidth]{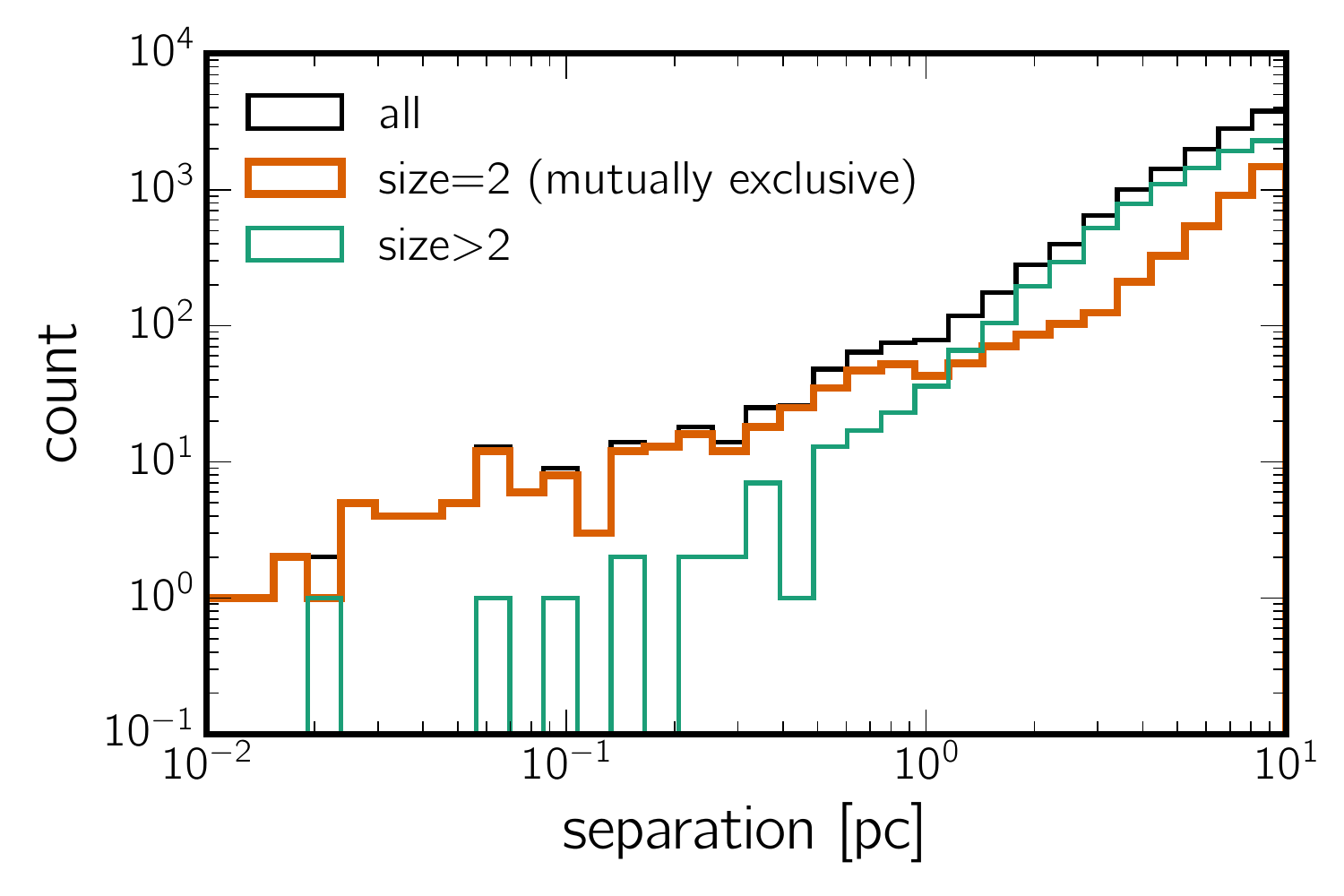}
  \end{center}
  \caption{%
    Separation distribution of comoving pairs of stars.
    As with Figure~\ref{fig:raverv},
    we divide the pairs into those mutually exclusively connected
    (connected component size $=2$), and those connected to larger connected components (size $>2$).
    \label{fig:hist_separation}
    }
\end{figure*}

We now examine the separation distribution of comoving pairs in
Figure~\ref{fig:hist_separation}.
As expected, pairs in larger connected components are mostly
found with separations larger than 1~pc.
Surprisingly, however, we also find a large number of mutually exclusive comoving
pairs
at $>1$~pc as well. Even if we consider the increasing false-positive rate at large
separations, the distribution is not significantly changed as the number of pairs
at $>1$~pc is in fact increasing much faster (as a power-law)
than the decrease due to the false-positives
(bottom panel of Figure~\ref{fig:raverv}).
The nature of these very wide separation, mutually exclusive pairs,
which cannot be gravitationally bound to each other, needs further investigation.
Can they be remnants of escaped binaries that are drifting apart?
In a study of the evolution of wide binaries including the Galactic tidal field
as well as passing field stars, \citet{Jiang:2010aa} found that we expect
to find a peak at $\sim 100-300$~pc in the projected separation due to
stars that were once in a wide binary system, but are drifting apart with small
relative velocities ($\sim 0.1$~\kms).
In future work, we will increase the maximum search limit (in this work, 10 pc)
to identify and study these large scale phase-space correlations.

\begin{figure*}[htbp]
  \begin{center}
    \includegraphics[width=\textwidth]{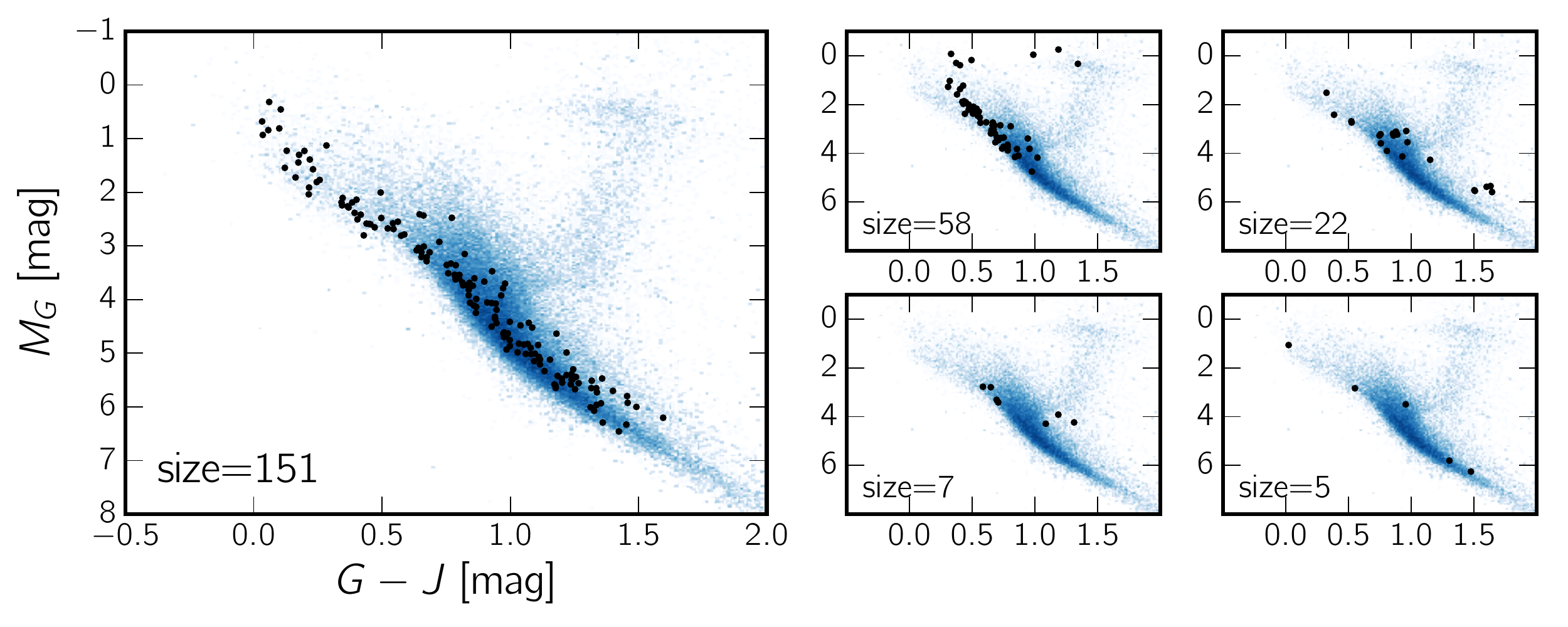}
  \end{center}
  \caption{
    Color-magnitude diagrams of stars in larger connected components presented in
    Figure~\ref{fig:graphviz_examples}. Each panel shows the same group
    of stars visualized in the panel at the same position
    in Figure~\ref{fig:graphviz_examples}.
    We show a reference distribution of stars randomly drawn to have
    a matching distribution in distance as the stars in our comoving pairs
    in blue.
    \label{fig:cmd_large}}
\end{figure*}

\begin{figure*}[htbp]
  \begin{center}
    \includegraphics[width=\textwidth]{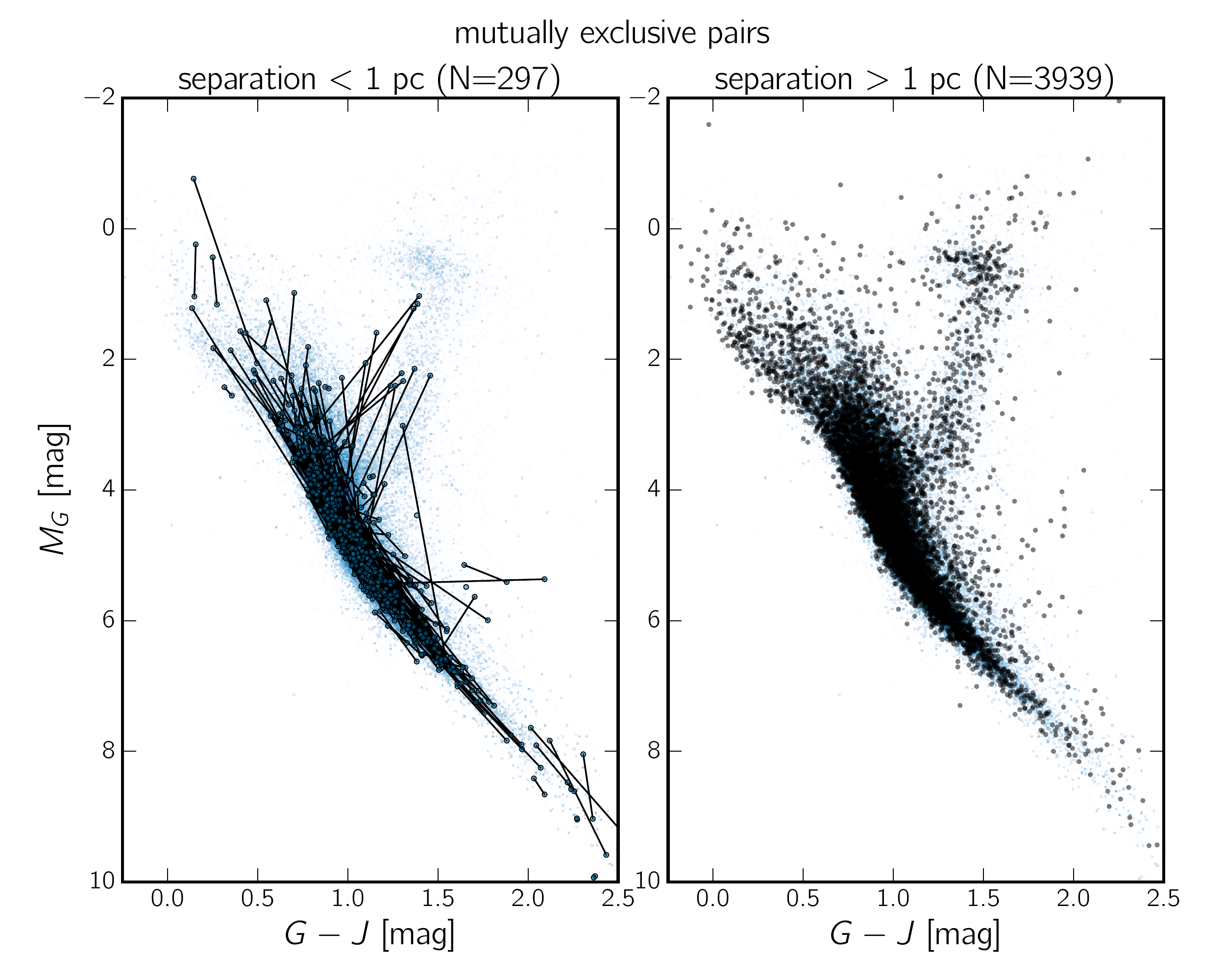}
  \end{center}
  \caption{
    Color-magnitude diagrams of stars in mutually exclusive comoving pairs
    in two separation bins.
    We connect each pair by a line on the left for pairs
    with separations smaller than 1~pc.
    We show a reference distribution of stars randomly drawn to have
    a matching distribution in distance as the stars in our comoving pairs
    in blue, same as Figure~\ref{fig:cmd_large}.
    \label{fig:cmd_me}}
\end{figure*}

Finally, we present the color-magnitude diagrams of comoving pairs
using the cross-matches with \tmass.
A more detailed study of stellar parameters using photometry from various sources
will follow. Figure~\ref{fig:cmd_large} and \ref{fig:cmd_me} shows $G-J$ vs $G$
color-magnitude diagrams for stars in larger connected components (size$>2$)
and in mutually exclusive pairs, respectively.
The connected components shown in Figure~\ref{fig:cmd_large}
correspond to those visualized
in Figure~\ref{fig:graphviz_examples}.
For stars in larger comoving groups, there is a noticeable
lack of evolved, off-main sequence
stars, in agreement with these kinematic structures being young.
For mutually exclusive pairs, we divide the pairs by separation at 1~pc
above which the false-positive rate due to random pairs starts to increase.
While many pairs are located along the main sequence, we also find
quite a few of main sequence-red giant pairs,
which will be valuable to anchoring stellar atmospheric models together.

\subsection{Catalog of candidate comoving pairs}
\label{sub:catalog}

\begin{figure}[htbp]
  \centering
  \includegraphics{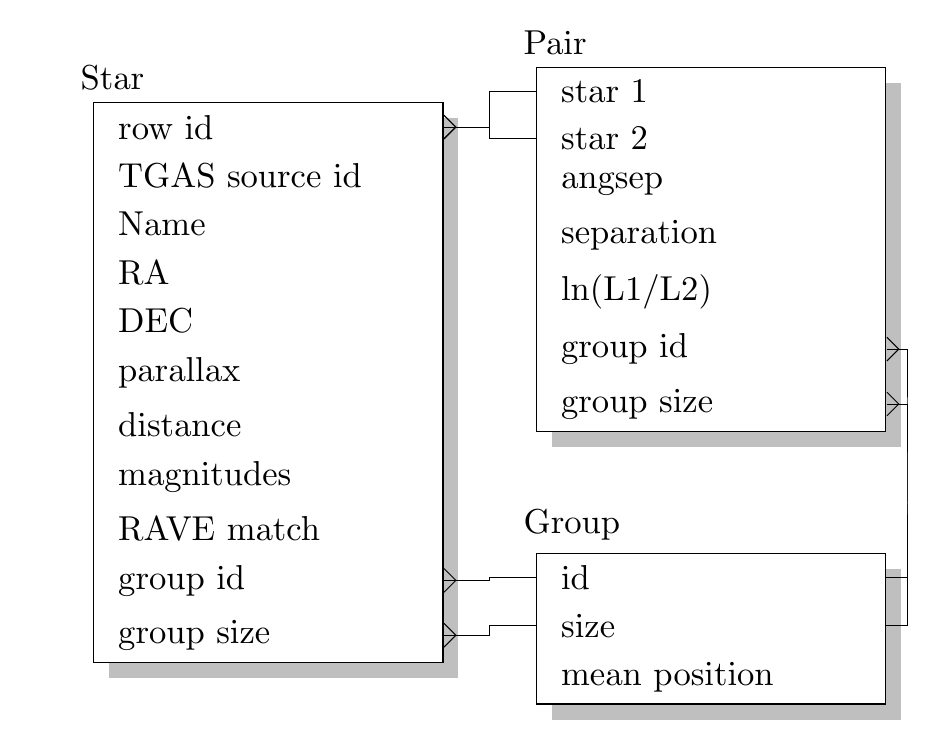}
  \caption{Schematic diagram of the relationship between the tables.
    See \sectionname~\ref{sub:catalog} and Table~\ref{tab:table_meta} for details.
    \label{fig:schema}}
\end{figure}

\begin{table}[htb]
\centering
\caption{Candidate co-moving pairs catalog description} \label{tab:table_meta}
\begin{tabular}{l|l|l}
\hline\hline
Column Name    & Unit & Description                         \\
\hline
\multicolumn{3}{c}{Table: Star (10,606 rows)}                                         \\
\hline
row id         &      & Zero-based row index                      \\
TGAS source id &      & Unique source id from \tgas               \\
Name           &      & \project{Hipparcos} or \project{Tycho-2} identifier \\
RA             & deg  & Right ascension from \tgas                \\
DEC            & deg  & Declination from \tgas                    \\
parallax       & mas  & Unique source id from \tgas               \\
distance       & pc   & Unique source id from \tgas               \\
$G$            & mag  & \gaia\ $G$-band magnitudes                \\
$J$            & mag  & \tmass\ $J$-band magnitudes               \\
RAVE OBS ID    &      & Unique id of the \rave\ match             \\
RV             & \kms & Radial velocity from \rave\               \\
eRV            & \kms & Uncertainty of radial velocity from \rave \\
group id       &      & Id of the group this star belongs to      \\
group size     &      & Size of the group this star belongs to    \\
\hline
\multicolumn{3}{c}{Table: Pair (13,058 rows)}                                            \\
\hline
star 1                           &        & Index of star 1 in the star table   \\
star 2                           &        & Index of star 2 in the star table   \\
angsep                           & arcmin & Angular separation                  \\
separation                       & pc     & Physical separation                 \\
$\ln\mathcal{L}_1/\mathcal{L}_2$ &        & Likelihood ratio                    \\
group id                         &        & Id of the group the pair belongs to \\
group size                       &        & Size of the group the pair belongs to \\
\hline
\multicolumn{3}{c}{Table: Group (4,555 rows)}                                         \\
\hline
id            &        & Unique group id                     \\
size          &        & Number of stars in a group          \\
mean RA       &  deg   & Mean right ascension of members     \\
mean DEC      &  deg   & Mean declination of members     \\
mean distance &  pc    & Mean distance of members     \\
\hline\hline
\end{tabular}
\end{table}

In this section, we describe our catalog of candidate comoving pairs of stars.
The catalog is composed of three tables of stars, pairs, and groups.
We summarize the content of each table in Table~\ref{tab:table_meta}, and
the relationships between the tables in Figure~\ref{fig:schema}.
The star table contains all 10,606 stars that have at least one comoving neighbor
by our selection. We provide the TGAS source id, which may be used to easily
retrieve cross matches between \gaia\ and other surveys
using the \gaia\ data archive.
For each star, we also include the positional measurements from \tgas,
\gaia\ $G$-band, \tmass\ $J$-band magnitudes, and
\rave\ radial velocities where they exist.
The comoving relationship between the stars is described in the pair table.
We also list the angular and physical separation of each pair, and the likelihood ratio,
$\ln \mathcal{L}_1 /\mathcal{L}_2(>6)$ (see \eqname~\ref{eq:hyp1} and \ref{eq:hyp2})
computed in this work.

Finally, the information about the connected components found in these comoving
star pairs is in the group table. We assign a unique index to each group
in descending order of its size. Thus, group 0 is the largest group
that contains 151 stars.
Each star or pair is associated with a group that it is a member of, listed
in the group id column of the star and pair table.

We note a caveat on the completeness of a connected component of comoving
pairs found in our catalog. Because we applied a simple cut in the likelihood
ratio ($\ln \mathcal{L}_1 /\mathcal{L}_2>6$), there is a possibility that, for
example, a star in a mutually exclusive pair in our catalog may still have
another possibly comoving companion which has been dropped because the
likelihood ratio is slightly less than 6.

\section{Summary}\label{sec:summary}

In this \documentname, we searched for comoving pairs of stars in the
\tgas\ catalog released as part of the \gaia\ DR1.
Our method is to compare the fully marginalized likelihoods between two
hypotheses: (1) that a pair of
stars shares the same 3D velocity, and (2) that the two stars have independent
3D velocities, in both cases incorporating the covariances of parallax and
proper motions.
We argued for a reasonable cut of the likelihood ratio, and found
13,058 candidate comoving pairs among 10,606 stars
with separations ranging from 0.005~pc to 10~pc, the limit of our search.

We found that some comoving pairs that we have identified are connected
by sharing a common comoving neighbor.
This network of comoving pairs, which forms an undirected graph,
can be decomposed into connected components in which any two stars are connected
by a path.
The entire 13,058 candidate comoving pairs are grouped into 4,555 connected
components. The most common is a size-2 connected component, i.e., the two
stars in these pairs are mutually exclusively linked.
Many of the larger connected components
naturally correspond to some of the known comoving structures
such as open clusters and stellar associations.
Some of these comoving groups of stars are newly discovered.

We have also found a large number of very wide separation ($>1$~pc)
mutually exclusive comoving pairs, in which the stars are the only comoving neighbor
of each other and not part of large connected components.
These are most likely remnants of dissolving wide binaries (\citealt{Jiang:2010aa}).
The abundance of highly probable wide separation comoving pairs
conclusively shows that there is no strict cut-off semi-major axis for wide binary systems
(e.g., \citealt{Wasserman:1987aa}).
The presence of these pairs and similar separation distribution have already
been noticed by \citealt{Shaya:2011aa} using the \project{Hipparcos} data.
If confirmed with radial velocity measurements, this population should still be
relatively young compared to the general disk field population.
Modeling the color-magnitude diagram distribution of these stars can shed some
light on this issue.
If they are remnants of dissolving systems that were born coeval,
the sample of very wide separation comoving pairs can potentially be used
to measure the recent ($\lesssim 1$~Gyr)
star formation history in the Solar neighborhood.
Comoving stars with separation less than 1~pc are very promising candidates
for wide binaries. They are found to be pairs of stars of varying stellar types.
Some of these pairs, such as main sequence-red giant or F/G/K-M dwarfs,
will be particularly valuable for testing theoretical
stellar models and calibrating observational measurements of low mass stars.

We note that a similar search for wide binaries using the \tgas\ data is performed
in a recent work by \citet{2016arXiv161107883O}. They find $\approx$1,900 wide binaries
with separation typically less than 1.5~pc, and 256 pairs with separation larger than
$\sim 1$~pc.
We emphasize that our method is based on a probabilistic model for the assumptions
on the \emph{3D velocities} of the two stars in a pair, and that we
marginalize over the (unknown) true distances and velocities of the stars
in contrast to just applying a cut in the proper motion space.

Finally, we make our catalog of 13,058 candidate comoving pairs available to the
community.
What we find using the \tgas\ is only a taste of what we will discover with
the future releases of the \gaia\ mission.

\acknowledgements

We thank Scott Tremaine for useful discussions.
This project was developed in part at the 2016 NYC Gaia Sprint, hosted by the Center for
Computational Astrophysics at the Flatiron Institute in New York City. The
Flatiron Institute is supported by the Simons Foundation.

This work has made use of data from the European Space Agency (ESA) mission
{\it Gaia} (\url{http://www.cosmos.esa.int/gaia}), processed by the {\it Gaia}
Data Processing and Analysis Consortium (DPAC,
\url{http://www.cosmos.esa.int/web/gaia/dpac/consortium}). Funding for the DPAC
has been provided by national institutions, in particular the institutions
participating in the {\it Gaia} Multilateral Agreement.

This research was partially supported by the \acronym{NSF} (grants
  \acronym{IIS-1124794}, \acronym{AST-1312863}, \acronym{AST-1517237}),
  \acronym{NASA} (grant \acronym{NNX12AI50G}),
  and the Moore-Sloan Data Science Environment at \acronym{NYU}. The data
analysis presented in this article was partially performed on computational
resources supported by the Princeton Institute for Computational Science and
Engineering (PICSciE) and the Office of Information Technology's High
Performance Computing Center and Visualization Laboratory at Princeton
University.

\software{The code used in this project is available from
\url{https://github.com/smoh/gaia-comoving-stars} under the MIT open-source
software license.
This research additionally utilized:
    \texttt{Astropy} (\citealt{Astropy-Collaboration:2013}),
    \texttt{IPython} (\citealt{Perez:2007}),
    \texttt{matplotlib} (\citealt{Hunter:2007}),
    and \texttt{numpy} (\citealt{Van-der-Walt:2011}).}


\bibliographystyle{aasjournal}
\bibliography{refs}

\appendix

\section{Relevant properties of Gaussian integrals}
\label{sec:appendixA}

In what follows, all vectors are column vectors, unless we have transposed them.
A relevant exponential integral solution is
\begin{eqnarray}
  \ln\left[\int\exp(-\frac{1}{2}\,
    \transp{[\vec{x}-\vec{\nu}]} \,
    \inv{\mat{A}} \,
    [\vec{x}-\vec{\nu}] - \Delta) \, \dd \vec{x}\right]
  &=& +\frac{1}{2}\ln ||2\pi\,\mat{A}|| -\Delta
  \quad , \label{eq:gauss-int}
\end{eqnarray}
where $\vec{x}$ and $\vec{\nu}$ are $D$-dimensional vectors, $\mat{A}$ is a
positive definite matrix, $\Delta$ is a scalar, and the integral is over all of
$D$-dimensional $\vec{x}$-space.
To cast our problem in this form, we will need to complete the square of the
exponential argument.
If we equate
\begin{eqnarray}
  \frac{1}{2}\,\transp{[\vec{x}-\vec{\nu}]}\,\inv{\mat{A}}\,[\vec{x}-\vec{\nu}] + \Delta
  &=& \frac{1}{2}\,\transp{\vec{x}}\,\inv{\mat{A}}\,\vec{x} + \transp{\vec{x}}\,\mat{B}\,\vec{b} + C
  \quad ,
\end{eqnarray}
where $\mat{B}\,\vec{b}$ is a $D$-vector, and $C$ is a scalar, then we find
\begin{eqnarray}
  \vec{\nu} &=& -\mat{A}\,\mat{B}\,\vec{b} \label{eq:convnu}
  \\
  \Delta & = & C - \frac{1}{2}\,\transp{\vec{\nu}}\,\inv{\mat{A}}\,\vec{\nu} \label{eq:convDelta}
  \quad .
\end{eqnarray}
We will identify terms in our likelihood functions with $\mat{A}$,
$\mat{B}\,\vec{b}$, and $C$, convert to $\vec{\nu}$ and $\Delta$ and compute
the marginalized likelihood using \eqname~\ref{eq:gauss-int}.

\section{Expressions for the marginalized likelihoods}\label{sec:appendix}

At given distance $r$, the velocity-marginalized likelihood can be computed
analytically using the expressions in Appendix~\ref{sec:appendixA}.
We will start by writing down expressions for the the likelihood multiplied by
the prior pdf for the velocities.
The likelihood for the data (proper motions of the two stars in a pair)
is a Gaussian (\eqname~\ref{eq:likefn}).
In order to simplify our notation,
we construct a velocity-space data vector $\vec{y}$ as follows:
\begin{equation}
  \vec{y} =
    \transp{\left(
      \begin{array}{c c c c}
        r_i\,\mu_{\alpha,i}^* &
        r_i\,\mu_{\delta,i} &
        r_j\,\mu_{\alpha,j}^* &
        r_j\,\mu_{\delta,j}
      \end{array}
    \right)}
\end{equation}
where the subscripts $i,j$ refer to the indices of each star in the pair
and we have multiplied the observables (the proper motions) by the distances
$r_i, r_j$, which is permitted because we are conditioning on the distances.
Fundamentally, our hypothesis 1 model (the stars have the same velocity with a
small difference) is
\begin{equation}
  \vec{y} = \mat{M} \, \vec{v} + \mathrm{noise}
\end{equation}
where now the $4 \times 3$ transformation matrix $\mat{M}$ is a stack of the
transformation matrices for each star computed from the pair of sky positions
and using \eqname~\ref{eq:transformation}.
The noise (in $\vec{y}$) is drawn from a $4 \times 4$ Gaussian with
block-diagonal covariance matrix, $\mat{\Sigma}$, constructed from
the proper motion covariance matrix of each star,
$\mat{C}_i, \mat{C}_j$, and the distances
$r_i, r_j$:
\begin{equation}
  \mat{\Sigma} = \left(
    \begin{array}{c c}
      r_i^2 \, \mat{C}_i & 0 \\
      0 & r_j^2 \, \mat{C}_j
    \end{array}
  \right)
\end{equation}

Given these definitions, the likelihood function for hypothesis 1 is
\begin{eqnarray}
  p(\data \given \vec{v}, r_i, r_j) &=& r_i^2\,r_j^2\,
    \normal(\vec{y} \given \mat{M}\,\vec{v}, \mat{\Sigma}) \\
  \ln p(\data \given \vec{v}, r_i, r_j) &=& 2\,\ln r_i + 2\,\ln r_j
    -\frac{1}{2}\,\ln||2\pi\,\mat{\Sigma}|| \nonumber \\
    && \quad -\frac{1}{2}\,\transp{[\vec{y}-\mat{M}\,\vec{v}]}\,
      \inv{\mat{\Sigma}}\,
      [\vec{y}-\mat{M}\,\vec{v}]
  \quad ,
\end{eqnarray}
where the factor of $r_i^2\,r_j^2$ is the Jacobian of the transformation from $\vec{y}$
to the data space ($\transp{(\mu_{\alpha,i}^*\ \,\mu_{\delta,i}\ \,\mu_{\alpha,j}^*\ \,\mu_{\delta,j})}$).

Now we multiply this likelihood with the velocity prior.
As described in \sectionname~\ref{sec:methods}
we use an isotropic, mixture-of-Gaussians prior on velocity (\eqname~\ref{eq:vprior}).
For simplicity here let us work out the marginalization for one component of
the mixture so that $\vec{v} \sim \normal(\vec{0}, \mat{V}_m)$.
Then,

\begin{eqnarray}
  \ln p(\vec{v}) = -\frac{1}{2}\,\ln||2\pi\,\mat{V}_m||
  -\frac{1}{2} \transp{\vec{v}}\,\inv{\mat{V}_m}\,\vec{v}
\end{eqnarray}
We can identify $\vec{\nu}$ and $\Delta$ in
$\ln p(\vec{v}) + \ln p(\data \given \vec{v}, r_i, r_j)$
using \eqname~\ref{eq:convnu} and \ref{eq:convDelta} as

\begin{eqnarray}
  \mat{A} &=& \inv{[\transp{\mat{M}}\,\inv{\mat{\Sigma}}\,\mat{M}+\inv{\mat{V}_m}]}
  \\
  \vec{\nu} &=& -\mat{A}\,\transp{\mat{M}}\,\inv{\mat{\Sigma}}\,\vec{y}
  \\
  \Delta &=& -2\,\ln r_i -2\,\ln r_j
    +\frac{1}{2}\,\ln||2\pi\,\mat{\Sigma}|| +\frac{1}{2}\,\ln||2\pi\,\mat{V}_m|| \nonumber \\
    && \quad +\frac{1}{2}\,\transp{\vec{y}}\,\inv{\mat{\Sigma}}\,\vec{y} -\frac{1}{2}\,\transp{\vec{\nu}}\,\inv{\mat{A}}\,\vec{\nu}
  \quad ,
\end{eqnarray}
which we plug in to \eqname~\ref{eq:gauss-int} to get the marginalized
likelihood conditioned on the two distances $r_i, r_j$.

The marginalized likelihood for the hypothesis 2 model (the stars have
independent velocities) is very similar.
In this case, the marginalized likelihood is a product of two independent
integrals $Q$, composed in the same way as the hypothesis 1 model but now for
each star individually, where
\begin{eqnarray}
  \vec{y} &=&
    \transp{\left(
      \begin{array}{c c}
        r\,\mu_{\alpha}^* & \quad
        r\,\mu_{\delta}
      \end{array}
    \right)}
  \\
  \mat{\Sigma} &=& r^2 \, \mat{C}
  \\
\end{eqnarray}
and $\mat{M}$ is now the transformation matrix for one star. Then,
\begin{eqnarray}
  \mat{A} &=& \inv{[\transp{\mat{M}}\,\inv{\mat{\Sigma}}\,\mat{M}+\inv{\mat{V}_m}]}
  \\
  \vec{\nu} &=& -\mat{A}\,\transp{\mat{M}}\,\inv{\mat{\Sigma}}\,\vec{y}
  \\
  \Delta &=& -2\,\ln r
    +\frac{1}{2}\,\ln||2\pi\,\mat{\Sigma}|| +\frac{1}{2}\,\ln||2\pi\,\mat{V}_m|| \nonumber \\
    && \quad +\frac{1}{2}\,\transp{\vec{y}}\,\inv{\mat{\Sigma}}\,\vec{y} -\frac{1}{2}\,\transp{\vec{\nu}}\,\inv{\mat{A}}\,\vec{\nu}
  \quad ,
\end{eqnarray}
and
\begin{equation}
  Q = \frac{1}{2}\ln ||2\pi\,\mat{A}|| -\Delta \quad .
\end{equation}

\end{document}